\documentclass[epjST]{svjour}
\usepackage{epsfig}

\usepackage{color}

\begin{document}

\title{Non-empirical pairing functional}
\author{%
T. Duguet\inst{1} \fnmsep\thanks{\email{duguet@nscl.msu.edu}} \fnmsep\thanks{Both authors wish to thank K.
Bennaceur for his collaboration on
            this on-going research program. This work was supported by the U.S.\
            National Science Foundation under Grant No.\ PHY-0456903.}
\and T. Lesinski\inst{2} \fnmsep\thanks{\email{lesinski@ipnl.in2p3.fr}}} \institute{National Superconducting
Cyclotron Laboratory and
           Department of Physics and Astronomy,
           Michigan State University, East Lansing, MI 48824, USA
           \and Universit{\'e} de Lyon, F-69003 Lyon, France;
           Institut de Physique Nucl{\'e}aire de Lyon,
           CNRS/IN2P3, Universit{\'e} Lyon 1, F-69622 Villeurbanne, France}
\abstract{The present contribution reports the first systematic finite-nucleus calculations performed using the
Energy Density Functional method and a {\it non-empirical} pairing functional derived from low-momentum
interactions. As a first step, the effects of Coulomb and the three-body force are omitted while only the bare
two-nucleon interaction at lowest order is considered. To cope with the finite-range and non-locality of the bare
nuclear interaction, the $^{1}S_{0}$ channel of $V_{\rm low \, k}$ is mapped onto a convenient operator form.
Neutron-neutron and proton-proton pairing correlations generated in finite nuclei by the direct term of the
two-nucleon interaction are characterized in a systematic manner. Eventually, such predictions are compared to
those obtained from empirical local functionals derived from density-dependent zero range interactions. The
characteristics of the latter are analyzed in view of that comparison and a specific modification of their
isovector density dependence is suggested to accommodate Coulomb effects and the isovector trend of neutron gaps
at the same time.}

 \maketitle

\section{Introduction}
\label{intro}

Low-Energy Nuclear Theory is currently going through an unprecedented revival. First, the explicit link between
Quantum Chromodynamics and inter-nucleon interactions is being realized through Effective Field Theory (EFT) based
on Chiral Perturbation Theory~\cite{epelbaum06a}. Second, the recent advent of low-momentum nuclear interactions
obtained through the application of Renormalization Group (RG) techniques~\cite{bogner03a} opens up, for the first
time, the possibility to understand properties of heavy nuclei from underlying microscopic interactions. This is
of crucial importance in view of the challenge posed by exotic nuclei displaying an unusually large ratio of
neutrons over protons. Indeed, many traditional features of nuclei close to the valley of stability tend to be
significantly modified as one adds more neutrons to the same element. As a matter of fact, the predictive power of
current theoretical methods is rather limited as one goes towards experimentally unknown nuclei. It is mandatory
to improve such a situation considering that ambitious experimental programs are being developed around the world
to synthesize and study medium-mass neutron-rich nuclei.

The nuclear Energy Density Functional (EDF) approach is the microscopic tool of choice to study medium-mass and
heavy nuclei in a systematic manner~\cite{bender03a}. Such an approach to finite nuclei strongly relies on the
concept of symmetry breaking and is formulated as a two-step method: (i) the Single-Reference (SR) formulation
that incorporates static collective correlations associated with symmetry-breaking modes (ii) the Multi-Reference
(MR) formulation\footnote{The Quasi-particle Random Phase Approximation is a limit case of the MR-EDF
approach~\cite{jancovici64a}.} that further includes quantum correlations associated with the fluctuations of the
phase and magnitude of the order parameters of broken symmetries~\cite{bender07a}. The EDF approach has had many
qualitative and quantitative successes over the last twenty years~\cite{bender03a} despite its empirical
construction by analogy to wave-function-based methods, i.e. Symmetry Unrestricted Hartree-Fock and the Generator
Coordinate methods. As recently discovered however, the empirical nature of the MR-EDF method leads to serious
pathologies that compromise its past and future predictions~\cite{doba07a}. Corrections to such pathologies are
currently being designed~\cite{lacroix07a}. The EDF method is also empirical in the sense that the functionals
used so far, e.g. Skyrme~\cite{vautherin72a} or Gogny~\cite{decharge80a} only have a very loose connection to
underlying inter-nucleon interactions. As a result, the predictive power of the method away from the regions where
the functional is constrained through the reproduction of selected experimental data is limited.

In the present contribution, we limit ourselves to the SR level and focus on static pairing correlations
incorporated through the breaking of $U(1)$ symmetry. Correlations associated with the fluctuations of the
phase~\cite{duguet03a} and the magnitude~\cite{meyer91a,bender06x} of the pairing gap can be further incorporated
at the MR level\footnote{Pairing correlations can also be grasped through symmetry conserving
approaches~\cite{volya01a,pillet02a,dussel07a}.}. Nucleonic pairing has a strong influence on all low-energy
properties of nuclei. This encompasses masses, separation energies, deformation, individual excitation spectra and
collective excitation modes such as rotations or vibrations. The role of pairing correlations is particularly
emphasized when going toward the neutron drip-line because of the proximity of the single-particle continuum.

As for the EDF as a whole, pairing functionals that have been used so far are of empirical character. Their
ability to reproduce pairing properties of nuclei close to the valley of stability, e.g. the Odd-Even Mass
Staggering (OEMS)~\cite{hilaire02a}, moment of inertia of rotational nuclei~\cite{bender03b} can be considered as
satisfactory. However, detailed and systematic characterizations of those observables and of individual excitation
spectra in even-even and odd-even nuclei as well as their dependence on the properties of the pairing functional
are still missing. In addition, there are hints from the spreading of the predictions that the predictive power of
existing empirical functionals as one goes to more neutron rich nuclei and enters the ``next major shell'' is very
limited~\cite{duguet05a}.

Our long-term objective is the construction of non-empirical energy density functionals derived explicitly from
inter-nucleon two-body and three-body interactions~\cite{duguet06a}. As already alluded to, the motivation to do
so lies in the fact that empirical EDFs leave unexplained all fitted data and, while they can make accurate
predictions for systems and properties that are sufficiently similar to those fitted, they can fail badly for
systems and properties that differ significantly from those fitted. Thanks to their suggested perturbative
nature~\cite{bogner05a}, low-momentum potentials (``$V_{\rm low \, k}$'')~\cite{bogner03a} offer the opportunity
to construct energy functionals from bare nuclear interactions for the first time\footnote{Here, a two-body
(three-body) {\it bare} interaction is in principle any interaction that fits the two-body (three-body) scattering
phase shifts in a low-energy domain that is physically relevant to low-energy nuclear structure and reproduce the
deuteron (triton and $^{3}He$) binding energy.}. As a very first step, we focus here on the pairing part of the
nuclear EDF.

The paper is organized as follows. Basic elements of the SR-EDF approach are recalled in section~\ref{formalism}.
In section~\ref{nonempirical}, the construction of the non-empirical pairing functional from low-momentum
interactions is briefly explained. The corresponding results are reported in section~\ref{results} for a set of
semi-magic nuclei and compared those obtained from traditional empirical pairing functionals. Our conclusions and
perspectives are given in section~\ref{conclu}.

\section{Elements of single-reference EDF formalism}
\label{formalism}

The SR-EDF including $U(1)$ symmetry breaking takes the form of a {\it generalized} Hartree-Fock-Bogoliubov (HFB)
formalism where the particle-hole and particle-particle parts of the EDF may resum correlations beyond the {\it
strict} Hartree, Fock and Bogoliubov diagrams calculated in terms of the bare nuclear Hamiltonian. The energy is
postulated under the form a functional $\mathcal{E}[\rho,\kappa,\kappa^{\ast}]$ of the (symmetry breaking)
one-body density $\rho$ and pairing tensor $\kappa$. The latter density matrices are mapped through an {\it
auxiliary} product state $|\Phi\rangle$,
\begin{equation}
\rho_{ij} \equiv \langle\Phi| \hat{a}^\dagger_j\,\hat{a}_i | \Phi\rangle
 \, , ~~~~~~~~~~~~~
\kappa_{ij} \equiv \langle\Phi| \hat{a}_j\,\hat{a}_i | \Phi\rangle \, ,
\end{equation}
where $\{\hat{a}^\dagger_i\}$ is an arbitrary single-particle basis. The minimization of
$\mathcal{E}[\rho,\kappa,\kappa^{\ast}]$ under particle-number constraints enforced through the use of a chemical
potential $\lambda$ leads to the generalized HFB equations
\begin{equation}
\left(
  \begin{array}{cc}
  h-\lambda   & \Delta  \\
-\Delta^{\ast}    &-h^{\ast}+\lambda
  \end{array} \right) \,
  \left(
  \begin{array}{c}
 \mathcal{U}  \\
\mathcal{V}
  \end{array} \right)_{\mu}
           = E_{\mu}  \,   \left(
  \begin{array}{c}
 \mathcal{U}  \\
\mathcal{V}
  \end{array} \right)_{\mu}
 \, \, \, , \label{HFBeigenvaluekappa}
\end{equation}
where $(\mathcal{U}, \mathcal{V})_{\mu}$ are the upper and lower components of the Bogoliubov quasi-particle
eigenstate whereas $E_{\mu}$ denotes the corresponding quasi-particle energy. Once the quasi-particle eigenstates
are obtained, $\rho$ and $\kappa$ can be recalculated and the iterative procedure processed until convergence. The
single-particle field $h$ and the pairing field $\Delta$ are obtained through functional derivatives
\begin{equation}
h_{ji} \equiv \frac{\delta \mathcal{E}}{\delta \rho_{ij}}
  =   h^{\ast}_{ij}, \, \, \,  \, \, \, \Delta_{ij}
  \equiv \frac{\delta \mathcal{E}}{\delta \kappa^{\ast}_{ij}}
    = - \Delta_{ji} \, \, \, . \label{fieldsdef}
\end{equation}

In a functional HFB scheme, the {\it particle-hole} and {\it particle-particle} channels can only be defined
rigourously at the level of the fields $h$ and $\Delta$, i.e. relatively to the density matrix with respect to
which the functional derivative is taken. Indeed, and except for the strict bilinear functional obtained from the
Hartree, Fock and Bogoliubov diagrams, the $\kappa$ and $\rho$ dependences are entangled in the energy
$\mathcal{E}[\rho,\kappa,\kappa^{\ast}]$ and one {\it cannot} in general split the functional into a particle-hole
and a particle-particle part. The common separation of the EDF into a particle-hole part on the one hand and a
pairing part on the other has only relied on the very simple functional dependence of standard phenomenological
functionals. Indeed, the only (bilinear) term depending on $\kappa$, possibly further depending on $\rho$, has
usually been characterized as the particle-particle part of the EDF, the rest then defining its particle-hole
part.

\section{Non-empirical pairing functional from low-momentum interactions}
\label{nonempirical}

\subsection{Constructive many-body framework}
\label{constructive1}

Instead of postulating the form of the nuclear EDF and fitting it to nuclear data, our long-term goal is to
construct it explicitly from many-body techniques formulated in terms of inter-nucleon interactions in the vacuum.
Starting from traditional two-body hard-core interactions, such an endeavor has always been perceived as
unrealistic because of the non-perturbative nature of the many-body problem and quantitatively unpractical because
of the Coester line problem~\cite{coester70a}. The advent of low-momentum interactions~\cite{bogner03a} and the
better understanding of the role played by three-nucleon forces make such a constructive approach promising for
the first time, at least in view of constraining the form and parameters of the nuclear EDF. As a matter of fact,
a SR many-body perturbation theory (MBPT) formulated in terms of low-momentum interactions becomes a viable
approach to building correlations into the functional~\cite{bogner05a}. This is the path we are going to follow.
Within such a scheme, the symmetry-breaking auxiliary state $|\Phi\rangle$ entering the SR-EDF formalism is
nothing but the vacuum on top of which the perturbation theory is performed.

As we treat pairing correlations through $U(1)$ symmetry breaking, MBPT must be formulated including anomalous
propagators (in terms of Feynman~\cite{gorkov58a} or Goldstone~\cite{henley64a} diagrams). At lowest order, the
irreducible vertex entering the pairing channel is given by the bare nuclear interaction ; i.e. it corresponds to
the strict Bogoliubov diagram. Omitting the three-nucleon interaction for now, the corresponding part of the EDF
is thus non local, bilinear in $\kappa$ and $\rho$-independent ; i.e. $\mathcal{E}[\rho,\kappa,\kappa^{\ast}]
\equiv \mathcal{E}[\rho]+\mathcal{E}^{\kappa\kappa}$, where $\mathcal{E}^{\kappa\kappa}$ denotes the bilinear
pairing functional. At the next order, the irreducible pairing vertex involves the so-called polarization
diagrams~\cite{gao06a,barranco04a}.

We limit ourselves in the present work to constructing the pairing part of the EDF from low-momentum interactions
and keeping the remaining part empirical, retaining only the lowest order contribution of the (nuclear part of)
the two-body interaction in the pairing energy. Of course, the full consistency of the approach will only be
attained when the Coulomb and three-nucleon forces are considered, when higher orders are incorporated and all
terms of the EDF are generated from underlying interactions. It is worth mentioning that the particle-hole part of
the EDF influences pairing properties since pair scattering strongly depends on the characteristics of the
single-particle field on top of which it develops. The latter developments are indeed envisioned within our long
term research program.

Despite the limitations of the first step undertaken in the present work, a significant amount of relevant physics
can be addressed already. For example, pairing correlations generated by the direct term of the two-nucleon
interaction have only been characterized so far in one finite nucleus, thanks to a very involved
calculation~\cite{barranco04a}. The contribution of the direct term of the two-nucleon interaction to the
magnitude and isotopic dependence of finite nuclei pairing gaps is however of general interest. Indeed, it is
unique to nuclear systems and trapped cold atoms that the direct term of the inter-particle interaction generates
superfluidity ; i.e. superconductivity in electronic systems is due to {\it induced} interactions generated by the
coupling to lattice phonons. By a straight comparison with experimental data, the characterization initiated in
this work will eventually suggest what is missing in the pairing functional beyond the direct term of the
two-nucleon interaction.

\subsection{Operatorial mapping of $V_{\rm low \, k}$}
\label{constructive2}

Limiting ourselves to the direct term of the bare two-nucleon interaction brings an extra simplification. As known
from scattering phase-shifts, only the $^{1}S_{0}$ channel of the nuclear interaction can generate pairing at
sub-saturation densities~\cite{dean03a}. As a result, all other partial waves can be omitted at this point. Of
course, this would not be true beyond lowest order and/or if constructing the particle-hole part of the EDF.

In the present work, the pairing functional is generated from $V_{\rm low \, k}$ calculated at a resolution
cut-off of $\Lambda \approx 2$ fm$^{-1}$. Traditionally, $V_{\rm low \, k}$ is generated through RG flow equations
and takes the form of tables of matrix elements in momentum space for each partial wave. For practical reasons, we
need an operator representation of the interaction which makes systematic EDF calculations of nuclei tractable. To
capture the finite range and non-locality of $V_{\rm low \, k}$ in a way that remains numerically tractable, we
produce a rank-$n$ separable representation of the interaction. Thus, focusing on the $^{1}S_{0}$ channel, the
spatial part of the interaction is represented as
\begin{eqnarray}
\langle \vec{r}_1 \vec{r}_2 \vert V^{S}_{\rm low \, k} \vert \vec{r}_3 \vec{r}_4 \rangle
    &\equiv& \sum_{\alpha,\beta=1}^{n}~
        G_\alpha(s_{12})~ \lambda_{\alpha\beta}~ G_\beta(s_{34})~
        \delta(\vec{R}_{12} - \vec{R}_{34})
 \, ,
\end{eqnarray}
where $\vec{R}_{12}=(\vec{r}_{1}+\vec{r}_{2})/2$ and $\vec{s}_{12}=\vec{r}_{1}-\vec{r}_{2}$ are the center-of-mass
and relative coordinates of the interacting-particle pair, respectively. The rank $n$, the form factors
$G_\alpha(r)$ and the coupling parameters $\lambda_{\alpha\beta}$ have to be specified and optimized. In fact, the
nuclear interaction is almost separable in the $^{1}S_{0}$ channel due to the presence of a virtual state at
almost zero scattering energy~\cite{brown}. As a result, a rank-$1$ representation is already quantitatively
satisfactory~\cite{duguet04a} and will be used in the present work\footnote{The low-energy phase shifts generated
by the rank-$1$ separable representation used better match experimental data than is apparent in Fig. 2 of
Ref~\cite{duguet04a}. This is because the results shown on that figure were plagued by a numerical error.}. Even
though the separable nature of the mapping brings noticeable simplifications~\cite{lesinski08a}, the finite-range
and non-local vertex cannot be handled easily in any existing HFB code. The high-precision operatorial mapping of
$V_{\rm low \, k}$ and the specificities of our new code~\cite{lesinski07a} will be discussed in detail in a
future work~\cite{lesinski08a}.

\section{First results in semi-magic nuclei}
\label{results}

As already mentioned, the entire EDF is not constructed consistently from low-momentum interactions at this point.
Thus, the non-empirical pairing functional is combined to a Skyrme EDF for the particle-hole part ; i.e. the SLy5
parametrization~\cite{chabanat98a}. The corresponding density-dependent and momentum-independent isoscalar
effective mass is equal to $m^{\ast}_{0}=0.7 \, m$ at saturation density. In section~\ref{subsec:3}, the
non-empirical pairing functional will be replaced by empirical ones derived from Density Dependent Delta
Interactions (DDDI). The calculations are performed with the BSLHFB code~\cite{lesinski07a} in a spherical box of
radius $R=20 \, fm$ and using a discretized Bessel basis $j_{l}(kr)$ with $k<k_{max}=4 \, fm^{-1}$ and a
partial-wave cutoff $j_{max}=45/2$. Using those ingredients, we calculate the properties of about 470 even-even
(predicted) spherical nuclei~\cite{database}.

In the present contribution, we limit ourselves to a single observable related to pairing correlations, i.e. the
odd-even mass staggering. In Ref.~\cite{lesinski08a}, the analysis will be deepened and several other observables
will be discussed. The connection between finite difference mass formulae employed to extract the OEMS and
theoretical gaps is less trivial than usually thought~\cite{duguet02b}. In the present case, experimental
three-point mass difference formula $\Delta^{(3)}_{q}(N/Z)$ centered on odd $N/Z$, will be compared to straight
theoretical gaps calculated in even-even nuclei. While this corresponds to the best zeroth-order comparison, a
more advanced treatment~\cite{duguet02b} will be employed in Ref.~\cite{lesinski08a} to reach precise quantitative
conclusions. The actual theoretical measures used are twofold (i) $\Delta^{q}_{LCS}$ is the canonical gap matrix
element associated with the lowest canonical quasi-particle energy (ii) $E^{q}_{LQP}$ is the lowest positive
quasi-particle energy solution of Eq.~\ref{HFBeigenvaluekappa}. As opposed to $\Delta^{q}_{LCS}$, $E^{q}_{LQP}$
also contains a contribution from the underlying single-particle spectrum. Where the two quantities significantly
differ corresponds to (sub-)shell closures which must be avoided in the discussion of static pairing correlations.

\subsection{Non-empirical pairing functional}
\label{subsec:2}

\begin{figure}
\centerline{\includegraphics[height=8.cm]{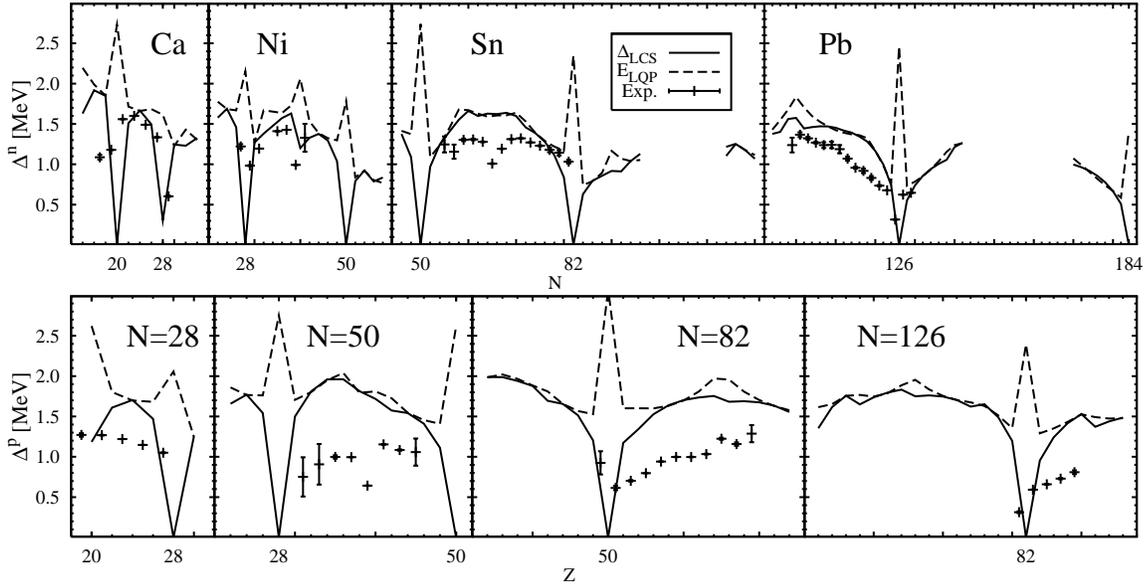}} \caption{Experimental three-point odd-even mass difference
(crosses) and zeroth-order theoretical pairing gaps obtained from the non-empirical pairing functional (lines).
Upper panel: neutron gaps along Ca, Ni, Sn and Pb isotopic chains. Lower panel: proton gaps along $N=28,50,82$ and
$126$ isotonic chains. Results are not displayed for nuclei predicted not to be spherical when our
criterion~\cite{rotival08a} is applied to the results of Ref.~\cite{database}.} \label{fig:1}
\end{figure}

The upper panel of Fig.~\ref{fig:1} displays the comparison between the experimental OEMS and theoretical neutron
gaps obtained from the non-empirical functional along isotopic chains of ``light'' (Ca and Ni), medium-mass (Sn)
and heavy (Pb) elements. One observes that (i) pairing gaps slightly decrease with increasing mass ; (ii) neutron
gaps obtained from the non-empirical functional are very close to experimental gaps in the lightest elements (iii)
but larger by a few hundreds keV in tin and lead isotopes ; (iv) theoretical neutron gaps, which cover a wider
range of isospin asymmetry than experimental data, display a decreasing slope with $N\!-\!Z$ along all isotopic
chains.

The agreement between experimental mass differences and the neutron gaps obtained from the non-empirical
functional is striking, to a large extent surprising, and to be taken with a grain of salt at this point. Indeed,
four crucial elements must be recalled at this point (i) gaps are exponentially sensitive to the interaction
strength ; (ii) the non-empirical functional is generated without {\it any} adjustment on finite-nucleus data
(iii) the non-empirical functional considered at this point is {\it not} final since it is only derived at lowest
order in the interaction ; i.e. it is not expected to reproduce experimental data, especially in view of the
recent claims that spin, isospin and density fluctuations beyond the direct term are expected to impact pairing
gaps significantly~\cite{barranco04a} ; (iv) the present results are based on an empirical particle-hole
functional which is not fully consistent with the pairing vertex. We will briefly come back to that below. In any
case, and even if the direct term provides neutron gaps of the right order of magnitude in nearly stable nuclei,
the three-body force and higher-order effects are likely to be crucial to reach a quantitative agreement with
experiment on a nucleus-by-nucleus basis as well as to understand isotopic trends as we go towards more
neutron-rich nuclei.

The lower panel of Fig.~\ref{fig:1} displays proton gaps along isotonic chains $N=28,50,82,126$. One observes that
experimental proton gaps decrease significantly with increasing mass and are, for a given mass region, smaller
than the neutron ones. One also observes that theoretical proton gaps obtained from the (isospin-invariant)
non-empirical pairing functional do not significantly decrease with mass, are rather flat as a function on
$N\!-\!Z$ and significantly overestimate experimental data, a discrepancy which increases with nuclear mass. The
difference with respect to the situation previously discussed for neutrons is striking. Qualitatively speaking,
the reduction of experimental proton gaps with respect to neutron ones is not visible in the theoretical results.
Of course, and even though other missing elements might be invoked, it is likely that the missing Coulomb
interaction is mainly responsible for that discrepancy. EDF calculations including Coulomb in the pairing have
been extremely rare~\cite{anguiano01a} and it will be of interest to test such a hypothesis~\cite{lesinski08a}.

It is of interest to compare $\Delta^{n}_{LCS}$ obtained in $^{120}$Sn to the one reported in
Ref.~\cite{barranco04a}, which is the only gap calculated in finite nuclei from a bare two-nucleon interaction
prior to the present work. The latter calculation was performed using the SLy4 Skyrme parametrization in the
particle-hole channel and the Argonne V18~\cite{wiringa95a} bare interaction in the pairing channel. It happens
that the predicted gap was about half of the one obtained here using the separable representation of $V_{\rm low
\, k}$\footnote{We do not quote precise values of the gaps because the calculation of Ref.~\cite{barranco04a} was
done using a reduced spin-orbit coupling strength as opposed to the original SLy4 parametrization. On the other
hand, it has been checked that the difference in the spin-orbit coupling is not sufficient to explain the
difference in the predicted gaps.}. It is surprising at first considering that both interactions provide identical
gaps in infinite nuclear matter when calculated with free single-particle energies~\cite{duguet04a}. The reason
for the difference seen in $^{120}$Sn relates to the {\it resolution scale} (the $\Lambda$ cut-off in RG terms) at
which the two-body interaction employed in the pairing channel and the effective mass characterizing the
single-particle field in the particle-hole channel are defined. Such a critical issue will be discussed at length
in a forthcoming publication~\cite{hebeler08a}. In any case, the relative agreement between experiment and the
gaps obtained presently, together with the qualitative discrepancy between the latter gaps and those displayed in
Ref.~\cite{barranco04a}, raise the question of the quantitative importance of higher-order effects associated with
the coupling to spin, isospin and density fluctuations.

\subsection{Comparison with empirical pairing functionals}
\label{subsec:3}

One of the goals of the present work is to corroborate the successes held by empirical local pairing functionals
and to help enriching them in a controlled manner. The latter objective is of particular interest considering that
the non-locality of the non-empirical functional makes it numerically demanding and difficult to adapt to 3D
codes. By a careful comparison of the results obtained from both approaches, one can thus hope to pin down
necessary isoscalar and isovector density dependences as well as genuine gradient corrections ; i.e. explicit
finite-range effects beyond the necessary ultra-violet regularization/renormalization of quasi-local pairing
functionals. Of course, the ``complete'' pairing functional will need to resum effects from higher-order
contributions and three-body forces and the comparison performed at this point can only account for lowest-order
effects from the two-body interaction. In addition, we only focus on the general trends of the OEMS and postpone a
detail analysis and the extension to other ground-state observables as well as excited states properties to a
future work~\cite{lesinski08a}.

For that purpose, we thus consider two empirical functionals derived from DDDI and regularized through a standard
cut-off scheme~\cite{doba84a} (i) REG-V is a density-independent local functional (ii) REG-S is a local functional
depending on the isoscalar density in such a way that its effective coupling strength is enhanced at the nuclear
surface (see Eq. 1 of Ref.~\cite{duguet05a}). The overall strength of both functionals is fixed to reproduce
$\Delta^{n}_{LCS}$ predicted in $^{120}$Sn by the non-empirical functional.

\begin{figure}
\centerline{\includegraphics[height=8.cm]{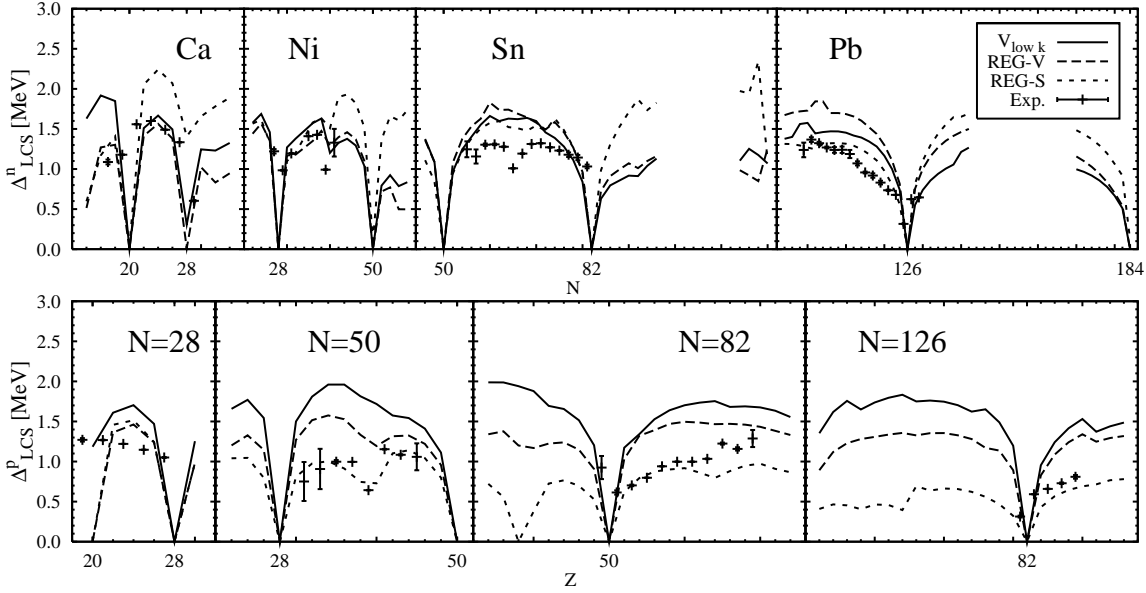}} \caption{Same as Fig.~\ref{fig:1} for gaps
obtained from empirical DDDI pairing functionals (see text).} \label{fig:2}
\end{figure}

Fig.~\ref{fig:2} adds to Fig.~\ref{fig:1} the gaps ($\Delta^{q}_{LCS}$) calculated from both empirical pairing
functionals. Let us first analyze the predicted neutrons gaps displayed in the upper panel of Fig.~\ref{fig:2}.
Keeping in mind that all neutron gaps are the same in $^{120}$Sn by construction, one observes, first, that the
gaps from REG-V are very close to those obtained with the non-empirical functional, i.e. the isotopic trend is
identical whereas the decrease with mass is only slightly less pronounced with REG-V. Moreover the isovector
character of REG-S is quite pronounced and opposite to the non-empirical functional. This can be linked to the
(surface-peaked) dependence of REG-S on the {\it isoscalar} density $\rho_{0}(\vec{r}) =
\rho_{n}(\vec{r})+\rho_{p}(\vec{r})$~\cite{duguet05b}. As opposed to a REG-S functional which would only depend on
$\rho_{n}(\vec{r})$, the dependence on $\rho_{0}(\vec{r})$ provides neutron gaps with a sharp increase as a
function of $N\!-\!Z$~\cite{duguet05b}. Although one cannot rule out that such an isovector dependence arises when
effects beyond the direct two-body interaction are included, the relative disagreement between REG-S and known
experimental data makes it rather suspicious at this point.

Let us now turn to proton gaps. One observes from the lower panel of Fig.~\ref{fig:2} that proton gaps from REG-V
and the non-empirical functional remain rather close, although the former are systematically smaller than the
latter. Proton gaps from REG-S are even smaller and decrease significantly with the nuclear mass, while slightly
increasing with the proton number. As a matter of fact, proton gaps from REG-S follow experimental data quite
closely. This is a non-trivial result considering, on the one hand, our conjecture that the qualitative difference
between experimental neutron and proton gaps is primarily due to Coulomb, and on the other hand that the REG-S
pairing functional is isospin invariant.

The qualitatively different behaviors of neutron and proton gaps as a function of isospin asymmetry and nuclear
mass can be traced back to the {\it surface-peaked} dependence of REG-S on the {\it isoscalar} density. In any
nucleus, the neutron and proton pairing fields obtained from REG-S are localized in the same region relative to
the surface because of the common density dependence on $\rho_{0}(\vec{r})$. However, in neutron-rich and/or
massive nuclei, where a neutron skin develops, the difference of radii of neutron and proton density distributions
makes the surface-peaked pairing field to overlap more with neutron orbitals than with proton orbitals. This
explains the strong increase of neutron gaps with neutron richness and reduction of proton gaps with nuclear mass
compared to those obtained from REG-V. Looking more carefully, one notices that the isovector dependence of
neutron pairing gaps predicted by REG-S is more prominent than for proton gaps. This calls for a more detailed
study of geometrical overlaps of densities, wave functions and density-dependent form factors, as well as an
investigation of the localizing effect of the Coulomb barrier on proton states~\cite{lesinski08a}. At this point
however, we may conjecture that an empirical REG-S functional depending only (mostly) on $\rho_{n}(\vec{r})$ could
provide a weaker isovector dependence of neutron gaps~\cite{duguet05b} on the one hand and mimic Coulomb effects
on proton gaps on the other hand\footnote{It would be at the price of dealing with an isospin-symmetry breaking
pairing functional.}. Such a conjecture needs to be tested thoroughly.

\section{Conclusions and outlook}
\label{conclu}

The first systematic finite-nucleus calculations performed using the Single-Reference Energy Density Functional
method and a non-empirical pairing functional derived from low-momentum interactions are reported in the present
work. As a first step, the effects of Coulomb and the three-body force on pairing are omitted while only the
direct term of the bare two-nucleon interaction is considered. Higher-order effects associated with the coupling
to spin, isospin and density fluctuations is postponed to later. The first step taking here towards the
construction of non-empirical energy density functionals already constitutes a challenge considering the
difficulty to treat the full finite-range and non-locality of the bare nuclear interaction in systematic Energy
Density Functional calculations. To do so, the $^{1}S_{0}$ channel of $V_{\rm low \, k}$ ($\Lambda \approx 2
fm^{-1}$) is mapped onto a convenient operator form, i.e. a (precise enough) rank-$1$ separable representation.

For the first time, pairing correlations generated in finite nuclei by the (lowest-order term of the) bare
two-nucleon interaction is characterized in a systematic manner. Restricting ourselves to one observable in the
present contribution, theoretical and experimental gaps are compared along several isotopic and isotonic chains of
semi-magic nuclei. The closeness of theoretical and experimental neutron gaps across several mass regions is
striking. Indeed, the non-empirical functional considered here is not final since it is only derived at lowest
order in the two-body interaction. On the other hand, theoretical proton gaps are systematically larger than
experimental ones which can be attributed to the omission of treating Coulomb in the pairing channel. It is one of
the goals of a forthcoming publication to prove this conjecture to be correct~\cite{lesinski08a}.

Eventually, predictions from the non-empirical functional are compared to those obtained from empirical local
functionals derived from density-dependent zero-range interactions. The characteristics of the latter are analyzed
in view of that comparison and a specific modification of their isovector density dependence is suggested to
accommodate Coulomb effects and the isovector trend of neutron gaps at the same time.

Beyond the results briefly summarized in the present contribution, it is the aim of a forthcoming publication to
(i) deepen and systematize the analysis, including a more advanced and quantitative evaluation of theoretical
pairing gaps (ii) discuss several other observables as well as the effect of Coulomb on proton pairing (iii) use a
higher-rank high-precision representation of $V_{\rm low \, k}$. Also, several extensions of the work presented
here are envisioned in the mid-term future. Beyond including the effect of Coulomb on pairing in a way that is
numerically tractable, our goal is to study the dependence of the results on the ``resolution scale'' ($\Lambda$
cut-off in renormalization group terms) at which the calculations are performed, approximate the non-empirical
functional through enriched quasi-local functionals, then use such approximations in codes working in a
three-dimensional representation as well as Multi-Reference Energy Density Functional calculations (in the sense
of the Generator Coordinate Method). Finally we could employ the latter formalism to study the coupling between
density, spin and isospin fluctuations and superfluidity. Ultimately, the effect of three-body forces on pairing
in finite nuclei should be investigated, yielding a consistent {\it ab-initio} picture.


\begin{thebibliography}{}
\bibitem{epelbaum06a}
E. Epelbaum, Prog. Part. Nucl. Phys. \textbf{57} (2006) 654
\bibitem{bogner03a}
S. K. Bogner, T. T. S. Kuo, A. Schwenk, Phys. Rept. \textbf{386} (2003) 1
\bibitem{bender03a}
M. Bender, P.-H. Heenen, P.-G. Reinhard, Rev. Mod. Phys. \textbf{75} (2003) 121
\bibitem{jancovici64a}
B. Jancovici, D. H. Schiff, Nucl. Phys. \textbf{58} (1964) 678
\bibitem{bender07a}
M. Bender, contribution to this volume
\bibitem{doba07a}
J. Dobaczewski, W. Nazarewicz, P. G. Reinhard, M. V. Stoitsov, unpublished, arXiv:0708.0441
\bibitem{lacroix07a}
D. Lacroix, T. Duguet, M. Bender, unpublished
\bibitem{vautherin72a}
D. Vautherin, D. M. Brink, Phys. Rev. \textbf{C5} (1972) 626
\bibitem{decharge80a}
J. Decharg\'e, D. Gogny, Phys. Rev. \textbf{C21} (1980) 1568
\bibitem{duguet03a}
T. Duguet, M. Bender, P. Bonche, P.-H. Heenen, Phys. Lett. \textbf{B559} (2003) 201
\bibitem{meyer91a}
J. Meyer, P. Bonche, J. Dobaczewski, H. Flocard, P. H. Heenen, Nucl. Phys. \textbf{A533} (1991) 307
\bibitem{bender06x}
M. Bender, T. Duguet, Int. J. Mod. Phys. \textbf{E16} (2007) 222
\bibitem{volya01a}
A. Volya, B. A. Brown, V. Zelevinsky, Phys. Lett. \textbf{B509} (2001) 37
\bibitem{pillet02a}
N. Pillet, P. Quentin, J. Libert, Nucl. Phys. \textbf{A697} (2002) 141
\bibitem{dussel07a}
G. G. Dussel, S. Pittel, J. Dukelsky, P. Sarriguren, Phys. Rev. \textbf{C76} (2007) 011302(R)
\bibitem{hilaire02a}
S. Hilaire, J.-F. Berger, M. Girod, W. Satula, P. Schuck, Phys. Lett. \textbf{B531} (2002) 61
\bibitem{bender03b}
M. Bender, P. Bonche, T. Duguet, P.-H. Heenen, Nucl. Phys. \textbf{A723} (2003) 354
\bibitem{duguet05a}
T. Duguet, K. Bennaceur, P. Bonche, YITP Report Series \textbf{112} (2005) B20
\bibitem{duguet06a}
T. Duguet, K. Bennaceur, T. Lesinski, J. Meyer, INT proceedings \textbf{15} (2006) 21
\bibitem{bogner05a}
S. K. Bogner, A. Schwenk, R. J. Furnstahl, A. Nogga, Nucl. Phys.  \textbf{A763} (2005) 59
\bibitem{coester70a}
F. Coester, S. Cohen, B. Day, C. M. Vincent, Phys. Rev. \textbf{C1} (1970) 769
\bibitem{gorkov58a}
L. P. Gorkov, Sov. Phys. JETP \textbf{34} (1958) 505
\bibitem{henley64a}
E. M. Henley and L. Wilets, Phys. Rev. \textbf{133} (1964) B1118
\bibitem{gao06a}
L. G. Cao, U. Lombardo, P. Schuck, Phys. Rev. \textbf{C74} (2006) 064301
\bibitem{barranco04a}
F. Barranco, R. A. Broglia, G. Colo', E. Vigezzi, P. F. Bortignon, Eur. Phys. J. \textbf{A21} (2004) 57
\bibitem{dean03a}
D. J. Dean, M. Hjorth-Jensen, Rev. Mod. Phys. \textbf{75} (2003) 607
\bibitem{lesinski08a}
T. Lesinski, T. Duguet, K. Bennaceur, unpublished
\bibitem{brown}
G. E. Brown, A. D. Jackson, {\it The Nucleon-Nucleon Interaction}, North-Holland, Amsterdam, 1976
\bibitem{duguet04a}
T. Duguet, Phys. Rev. \textbf{C69} (2004) 054317
\bibitem{lesinski07a}
T. Lesinski, unpublished
\bibitem{chabanat98a}
E. Chabanat, P. Bonche, P. Haensel, J. Meyer, R. Schaeffer, Nucl. Phys. \textbf{A635} (1998) 231
\bibitem{duguet02b}
T. Duguet and P. Bonche and P.-H. Heenen, J. Meyer, Phys. Rev. \textbf{C65} (2002) 014311
\bibitem{rotival08a}
V. Rotival, K. Bennaceur, T. Duguet, unpublished, arXiv:0711.1275
\bibitem{database}
S. Hilaire, M. Girod, Eur. Phys. J. \textbf{A33} (2007) 237
\bibitem{anguiano01a}
M. Anguiano, J. L. Egido, L. M. Robledo, Nucl. Phys. \textbf{A683} (2001) 227
\bibitem{wiringa95a}
R. B. Wiringa, V. G. J. Stoks, R. Schiavilla, Phys. Rev. \textbf{C51} (1995) 38
\bibitem{doba84a}
J. Dobaczewski, H. Flocard, J. Treiner, Nucl. Phys. \textbf{A422} (1984) 103
\bibitem{duguet05b}
T. Duguet, P. Bonche, AIP Conf. Proc. 764 (2005) 277
\bibitem{hebeler08a}
K. Hebeler, T. Lesinski, T. Duguet, A. Schwenk, S. K. Bogner, K. Bennaceur, unpublished
\end{thebibliography}
\end{document}